\documentclass[conference]{IEEEtran}
\IEEEoverridecommandlockouts

\usepackage{cite}
\usepackage{amsmath,amssymb,amsfonts}
\usepackage{algorithmic}
\usepackage{graphicx}
\usepackage{textcomp}
\usepackage{xcolor}
\usepackage{tikz}
\usetikzlibrary{arrows.meta,calc,matrix,positioning,fit,backgrounds}
\usetikzlibrary{arrows.meta,calc}
\def\BibTeX{{\rm B\kern-.05em{\sc i\kern-.025em b}\kern-.08em
T\kern-.1667em\lower.7ex\hbox{E}\kern-.125emX}}

\usepackage{pgfplots}
\usetikzlibrary{arrows.meta,decorations.pathreplacing}
\usepgfplotslibrary{groupplots}
\pgfplotsset{compat=1.18}

\makeatletter
\newcommand{\linebreakand}{%
\end{@IEEEauthorhalign}
\hfill\mbox{}\par
\mbox{}\hfill\begin{@IEEEauthorhalign}
}
\makeatother

\begin{document}
\raggedbottom
    \title{TREDD: \\ Robust Trend-Based Reference Evaluation for Interpretable Degradation Detection\\
    \thanks{This work was supported by the Federal Ministry of Transport (BMV) under research grant number 01FV2063A.}
    }

	\author{
		\IEEEauthorblockN{Elisabeth Vogel}
		\IEEEauthorblockA{\textit{Chair of Wireless Systems} \\
			\textit{Brandenburg University of Technology Cottbus-Senftenberg}\\
			Cottbus, Germany \\
			Elisabeth.Vogel@b-tu.de}
		\and
		\IEEEauthorblockN{Ronny Porsch}
		\IEEEauthorblockA{\textit{Chair of AI-powered Automation Technology} \\
			\textit{Brandenburg University of Technology Cottbus-Senftenberg}\\
			Cottbus, Germany \\
			Ronny.Porsch@b-tu.de}
		
		\linebreakand
		
		\IEEEauthorblockN{Peter Langendoerfer}
		\IEEEauthorblockA{\textit{Chair of Wireless Systems} \\
			\textit{Brandenburg University of Technology Cottbus-Senftenberg}\\
			Cottbus, Germany \\
			Peter.Langendoerfer@b-tu.de}
	}
	
	\maketitle

    \begin{abstract}
    	Technical systems are increasingly monitored using sensor and operational data to detect degradation and performance deterioration at an early stage. However, observed trends in measurement data do not necessarily correspond to physical aging, since noise, outliers, unstable initial regions, or changing operating conditions may produce similar patterns. This paper proposes Trend-based Reference Evaluation for Degradation Detection (TREDD), an interpretable method for detecting degradation as a persistent, trend-based deviation from an early reference state. TREDD combines rolling-window smoothing, baseline estimation, a direction-dependent degradation index, long-term trend extraction, and persistent drift detection. In addition, the method separates computational drift detection from the interpretation of drift as plausible physical aging by incorporating data-quality assessment, context checking, and robust auxiliary analysis. The approach is evaluated on the NASA Lithium-Ion Battery Aging Dataset using discharge-cycle capacity as the degradation-relevant condition variable. The representative case study illustrates that TREDD can identify clear degradation trajectories while assigning reduced confidence to weak, gradual, or atypical trends. The method therefore supports transparent and confidence-based degradation interpretation rather than purely predictive battery health estimation.
    \end{abstract}


    \section{Introduction}
    Technical systems are increasingly monitored using sensors and operational data to detect degradation, aging, or the onset of performance deterioration at an early stage. However, a key challenge is that observed changes in measurement data do not automatically correspond to physical degradation. Measurement noise, outliers, unstable initial conditions, or changing operating conditions can produce similar trend patterns, leading to unreliable interpretations.
    
    This paper therefore proposes Trend-based Reference Evaluation for Degradation Detection (TREDD). TREDD is an interpretable approach for detecting persistent, trend-based deviations of a condition-relevant variable from an early reference state. The approach combines rolling-window smoothing, baseline estimation, a direction-dependent degradation index, and the determination of drift start, degradation magnitude, and degradation rate. Additionally, TREDD distinguishes between computational drift detection and its interpretation as plausible physical aging.
    
    To support this interpretation, TREDD takes into account data quality, contextual information, and robust auxiliary analysis. As a result, weak, gradual, or atypical trends are not prematurely classified as definitive aging but are instead categorized with reduced confidence. The approach is evaluated on the publicly available NASA Lithium-Ion Battery Aging Dataset, using the capacity measured per discharge cycle as the degradation-relevant condition variable.
    
    The main contributions of this paper are as follows:
    \begin{itemize}
    	\item We introduce TREDD, an interpretable method for detecting degradation as a persistent trend-based deviation from an early reference state.
    	\item We separate computational drift detection from the interpretation of drift as plausible physical aging by incorporating data-quality assessment, context checking, and robust auxiliary analysis.
    	\item We evaluate TREDD on the NASA Lithium-Ion Battery Aging Dataset and show how the method distinguishes clear, weak, gradual, and atypical degradation trajectories with different confidence levels.
    \end{itemize}
    
    \section{Related Work}
    Although TREDD is formulated as a general method for degradation detection in monitored technical systems, the related work in this paper focuses on battery-health assessment because the experimental evaluation is conducted on lithium-ion battery aging data. The battery domain provides a suitable evaluation case, since discharge capacity is a physically meaningful condition variable and the NASA Lithium-Ion Battery Aging Dataset provides repeated degradation trajectories \cite{NASA_Battery_Dataset,Berecibar2016SOH}.
    
    Within prognostics and battery-health assessment, recent work is strongly dominated by data-driven approaches for remaining useful life (RUL), state of health (SOH), and state of charge (SOC) prediction.
	
	Gao, Aziz, and Hussain~\cite{b1} proposed a Three-Step Similarity Evaluation to identify batteries in a training set that are most similar to the monitored battery, thereby reducing prediction errors caused by manufacturing variations. Fan, Liang, and Zhang~\cite{b2} introduced a general RUL prediction framework based on a temporal convolutional network (TCN), where an adaptive consecutive validation mechanism is used to distinguish persistent degradation-relevant spikes from noise. Zhang et al.~\cite{b3} addressed the dependence of RUL prediction on large labeled datasets by combining a self-attention mechanism with a variational autoencoder (VAE) to learn from both labeled and unlabeled data.
	 
	Several approaches rely on deep neural architectures to model temporal degradation patterns. Shi et al.~\cite{b4} combined an encoder-decoder architecture with bidirectional gated recurrent unit (BiGRU) networks, where the encoder captures aging features and the decoder predicts RUL. The method was validated on the NASA battery dataset. Xu et al.~\cite{b5} also evaluated their approach on the NASA battery dataset and used a transformer-based convolutional neural network for noise reduction, followed by bidirectional modeling of temporal dependencies. 
	 
	A further line of work addresses interpretability in battery health prediction. Huang et al.~\cite{b7} combined feature engineering with model interpretability analysis to extract electrochemical-related features for random-forest-based SOH estimation. Farghaly et al.~\cite{b8} proposed an explainable SOH prediction approach on the NASA battery dataset by combining a transformer model with Shapley additive explanations (SHAP). Other studies jointly consider different battery health indicators. Indumathi and Gopalakrishnan~\cite{b9} argued that SOC and SOH are often treated separately despite their physical interdependence and proposed a transformer-based framework for jointly modeling short-term SOC dynamics and long-term SOH degradation. Guo et al.~\cite{b10} similarly estimated short-term SOC and long-term SOH using a hybrid temporal convolutional network and employed SHAP analysis to interpret degradation behavior.
	 
	In contrast to these prediction-oriented approaches, TREDD does not primarily aim to predict RUL, SOH, or SOC. Instead, it focuses on the interpretable detection and assessment of persistent degradation trends relative to an early reference state. The proposed method separates computational drift detection from the interpretation of drift as plausible physical aging by incorporating data-quality assessment, context checking, and robust auxiliary analysis. This enables TREDD to distinguish clear degradation trajectories from weak, gradual, or atypical trends with different confidence levels.
    
    \section{TREDD}
    \label{sec:tredd}
    This section describes the proposed TREDD approach. The aim of the method is to detect degradation as a persistent, trend-based deviation of a condition-relevant variable from an early reference state.
    
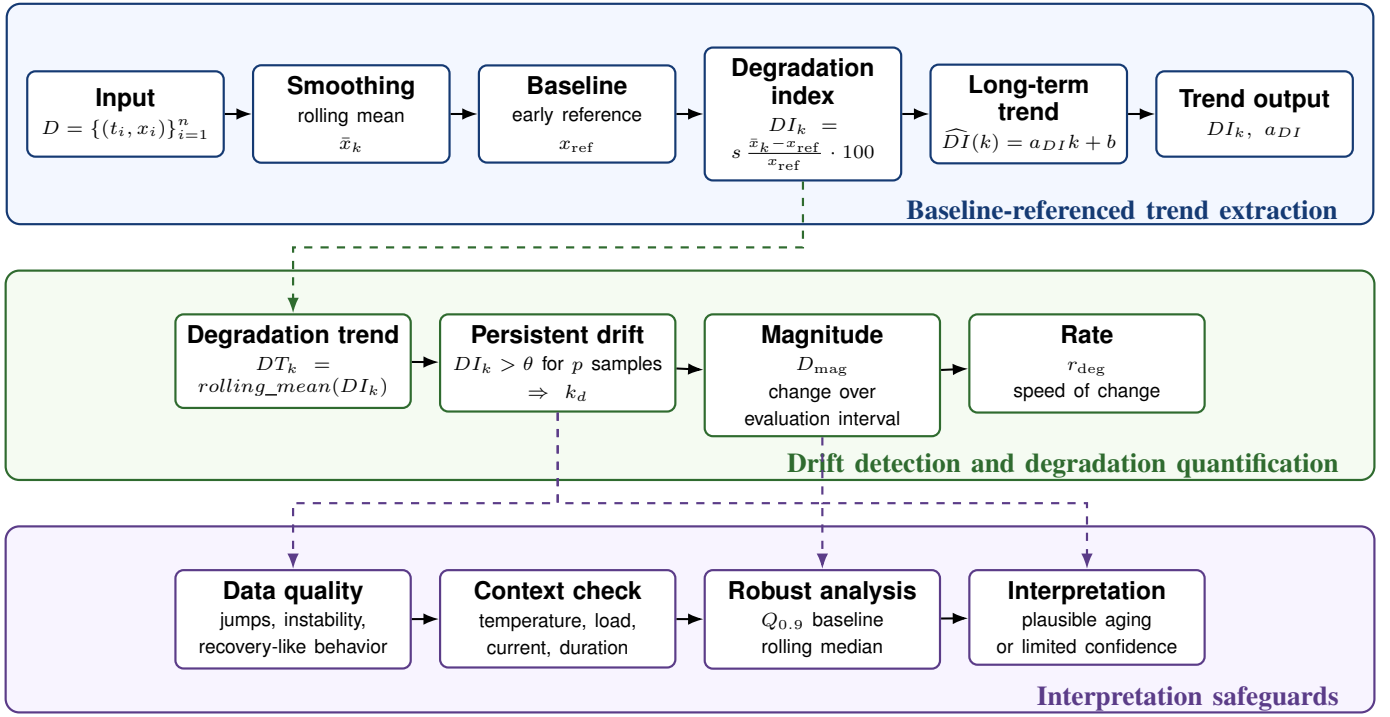
\begin{figure*}[t]
    	\centering
    	\resizebox{\textwidth}{!}{%
    		\begin{tikzpicture}[
    			font=\sffamily\small,
    			>=Latex,
    			proc/.style={
    				draw=#1,
    				rounded corners=3pt,
    				line width=0.8pt,
    				fill=white,
    				align=center,
    				inner sep=4pt,
    				minimum height=1.10cm,
    				text width=2.25cm
    			},
    			wideproc/.style={
    				draw=#1,
    				rounded corners=3pt,
    				line width=0.8pt,
    				fill=white,
    				align=center,
    				inner sep=4pt,
    				minimum height=1.10cm,
    				text width=2.75cm
    			},
    			arr/.style={
    				-{Latex[length=2mm]},
    				line width=0.75pt
    			},
    			darrgreen/.style={
    				dashed,
    				draw=treddgreen,
    				-{Latex[length=2mm]},
    				line width=0.8pt
    			},
    			darrpurple/.style={
    				dashed,
    				draw=treddpurple,
    				-{Latex[length=2mm]},
    				line width=0.8pt
    			}
    			]
    			
    			\definecolor{treddblue}{HTML}{153A73}
    			\definecolor{treddgreen}{HTML}{2F6B2F}
    			\definecolor{treddpurple}{HTML}{5B3C88}
    			\definecolor{treddlightblue}{HTML}{F3F7FF}
    			\definecolor{treddlightgreen}{HTML}{F5FBF2}
    			\definecolor{treddlightpurple}{HTML}{F8F4FF}
    			
    			\matrix (Arow) [
    			matrix,
    			column sep=0.35cm,
    			ampersand replacement=\&
    			] {
    				\node[proc=treddblue] (input)
    				{\textbf{Input}\\
    					{\scriptsize \(D=\{(t_i,x_i)\}_{i=1}^{n}\)}}; \&
    				
    				\node[proc=treddblue] (smooth)
    				{\textbf{Smoothing}\\
    					{\scriptsize rolling mean}\\
    					{\scriptsize \(\bar{x}_k\)}}; \&
    				
    				\node[proc=treddblue] (base)
    				{\textbf{Baseline}\\
    					{\scriptsize early reference}\\
    					{\scriptsize \(x_{\mathrm{ref}}\)}}; \&
    				
    				\node[proc=treddblue] (di)
    				{\textbf{Degradation}\\
    					\textbf{index}\\
    					{\scriptsize \(DI_k =
    						s\,\frac{\bar{x}_k-x_{\mathrm{ref}}}
    						{x_{\mathrm{ref}}}\cdot100\)}}; \&
    				
    				\node[proc=treddblue] (trend)
    				{\textbf{Long-term}\\
    					\textbf{trend}\\
    					{\scriptsize \(\widehat{DI}(k)=a_{DI}k+b\)}}; \&
    				
    				\node[proc=treddblue] (tout)
    				{\textbf{Trend output}\\
    					{\scriptsize \(DI_k,\ a_{DI}\)}}; \\
    			};
    			
    			\matrix (Brow) [
    			matrix,
    			below=1.45cm of Arow,
    			column sep=0.35cm,
    			ampersand replacement=\&
    			] {
    				\node[wideproc=treddgreen] (dt)
    				{\textbf{Degradation trend}\\
    					{\scriptsize \(DT_k =\)}\\[-2pt]
    					{\scriptsize \(rolling\_mean(DI_k)\)}}; \&
    				
    				\node[wideproc=treddgreen] (drift)
    				{\textbf{Persistent drift}\\
    					{\scriptsize \(DI_k > \theta\) for \(p\) samples}\\
    					{\scriptsize \(\Rightarrow k_d\)}}; \&
    				
    				\node[wideproc=treddgreen] (mag)
    				{\textbf{Magnitude}\\
    					{\scriptsize \(D_{\mathrm{mag}}\)}\\
    					{\scriptsize change over evaluation interval}}; \&
    				
    				\node[wideproc=treddgreen] (rate)
    				{\textbf{Rate}\\
    					{\scriptsize \(r_{\mathrm{deg}}\)}\\
    					{\scriptsize speed of change}}; \\
    			};
    			
    			\matrix (Crow) [
    			matrix,
    			below=1.45cm of Brow,
    			column sep=0.35cm,
    			ampersand replacement=\&
    			] {
    				\node[wideproc=treddpurple] (quality)
    				{\textbf{Data quality}\\
    					{\scriptsize jumps, instability,}\\
    					{\scriptsize recovery-like behavior}}; \&
    				
    				\node[wideproc=treddpurple] (context)
    				{\textbf{Context check}\\
    					{\scriptsize temperature, load,}\\
    					{\scriptsize current, duration}}; \&
    				
    				\node[wideproc=treddpurple] (robust)
    				{\textbf{Robust analysis}\\
    					{\scriptsize \(Q_{0.9}\) baseline}\\
    					{\scriptsize rolling median}}; \&
    				
    				\node[wideproc=treddpurple] (final)
    				{\textbf{Interpretation}\\
    					{\scriptsize plausible aging}\\
    					{\scriptsize or limited confidence}}; \\
    			};
    			
    			\begin{scope}[on background layer]
    				
    				\node[
    				draw=treddblue,
    				fill=treddlightblue,
    				rounded corners=5pt,
    				line width=0.8pt,
    				fit=(input)(smooth)(base)(di)(trend)(tout),
    				inner xsep=0.25cm,
    				inner ysep=0.55cm
    				] (Abg) {};
    				
    				\coordinate (Bleft)  at (Abg.west |- dt.center);
    				\coordinate (Bright) at (Abg.east |- dt.center);
    				
    				\coordinate (Cleft)  at (Abg.west |- quality.center);
    				\coordinate (Cright) at (Abg.east |- quality.center);
    				
    				\node[
    				draw=treddgreen,
    				fill=treddlightgreen,
    				rounded corners=5pt,
    				line width=0.8pt,
    				fit=(Bleft)(Bright)(dt)(drift)(mag)(rate),
    				inner xsep=0.00cm,
    				inner ysep=0.55cm
    				] (Bbg) {};
    				
    				\node[
    				draw=treddpurple,
    				fill=treddlightpurple,
    				rounded corners=5pt,
    				line width=0.8pt,
    				fit=(Cleft)(Cright)(quality)(context)(robust)(final),
    				inner xsep=0.00cm,
    				inner ysep=0.55cm
    				] (Cbg) {};
    				
    			\end{scope}
    			
    			\node[anchor=east, text=treddblue, font=\bfseries]
    			at ([xshift=-0.35cm,yshift=0.20cm]Abg.south east)
    			{Baseline-referenced trend extraction};
    			
    			\node[anchor=east, text=treddgreen, font=\bfseries]
    			at ([xshift=-0.35cm,yshift=0.20cm]Bbg.south east)
    			{Drift detection and degradation quantification};
    			
    			\node[anchor=east, text=treddpurple, font=\bfseries]
    			at ([xshift=-0.35cm,yshift=0.20cm]Cbg.south east)
    			{Interpretation safeguards};
    			
    			\draw[arr] (input) -- (smooth);
    			\draw[arr] (smooth) -- (base);
    			\draw[arr] (base) -- (di);
    			\draw[arr] (di) -- (trend);
    			\draw[arr] (trend) -- (tout);
    			
    			\draw[arr] (dt) -- (drift);
    			\draw[arr] (drift) -- (mag);
    			\draw[arr] (mag) -- (rate);
    			
    			\draw[arr] (quality) -- (context);
    			\draw[arr] (context) -- (robust);
    			\draw[arr] (robust) -- (final);
    			
    			
    			\coordinate (ABgap) at ($(Abg.south)!0.5!(Bbg.north)$);
    			\coordinate (BCgap) at ($(Bbg.south)!0.5!(Cbg.north)$);
    			
    			\draw[darrgreen]
    			(di.south) --
    			(di.south |- ABgap) --
    			(dt.north |- ABgap) --
    			(dt.north);
    			
    			\draw[darrpurple]
    			(drift.south) --
    			(drift.south |- BCgap) --
    			(quality.north |- BCgap) --
    			(quality.north);
    			
    			\draw[darrpurple]
    			(drift.south) --
    			(drift.south |- BCgap) --
    			(robust.north |- BCgap) --
    			(robust.north);
    			
    			\draw[darrpurple]
    			(mag.south) --
    			(mag.south |- BCgap) --
    			(final.north |- BCgap) --
    			(final.north);
    			
    		\end{tikzpicture}%
    	}
    	\caption{Overview of the complete TREDD pipeline. The first two levels form the computational core of the method, while the third level provides interpretation safeguards through data-quality assessment, context checking, and robust auxiliary analysis.}
    	\label{fig:tredd_pipeline}
    \end{figure*}

    The approach is formulated in general terms for technical systems that can be monitored through a temporally or cyclically ordered condition variable. In the case of battery aging, this variable is the capacity measured in each discharge cycle. In principle, however, the same methodology can also be applied to other degradation-relevant variables, such as power, efficiency, thermal behavior, pressure, or vibration features. Fig.  \ref{fig:tredd_pipeline} shows the overall structure of the TREDD pipeline.
    
    The first two levels (blue and green) form the computational core of the method and are described in this section. They cover the steps from the initial measurement series through smoothing, baseline estimation, degradation index calculation, and long-term trend extraction to drift detection and the calculation of degradation magnitude and degradation rate. The third level (purple) provides interpretation safeguards through data-quality assessment, context checking, and robust auxiliary analysis, and is explained in Section~\ref{sec:robustness}.
	
	\subsection{Input Data and Temporal Order}
	The starting point is a measurement series%
	\footnote{Unless stated otherwise, mathematical symbols are used in their standard meaning throughout this paper. In particular, \(\mu\) denotes a mean value and \(\sigma\) denotes a standard deviation.}
	\begin{equation}
		D = \{(t_i, x_i)\}_{i=1}^{n}
	\end{equation}
	where \(t_i\) denotes the time, cycle, or observation index, and \(x_i\) represents a condition-relevant measured variable. For the application to lithium-ion batteries, \(x_i\) corresponds to the capacity measured in discharge cycle \(i\).
	
	Since degradation is a time-dependent process, the measurement series must be arranged in a consistent temporal order. It is therefore ensured that
	\begin{equation}
	t_1 \leq t_2 \leq \dots \leq t_n 
	\label{eq2}
	\end{equation}
	holds. The series is then considered as the ordered sequence
	\begin{equation}
	x_1, x_2, \dots, x_n. 
	\label{eq3}
	\end{equation}
	This step is necessary because all subsequent calculations, in particular smoothing, trend analysis, and drift detection, depend directly on the correct temporal order.
	
	\subsection{Rolling-Window Smoothing}
	Real-world measurement series often contain short-term fluctuations, measurement noise, or individual outliers. To avoid reacting to individual measurement points, the condition variable is first smoothed using a rolling-window method. The smoothed value \(\bar{x}_k\) is calculated as
	\begin{equation}
		\bar{x}_k =
		\frac{1}{w_k}
		\sum_{j=\max\{1,k-w+1\}}^{k} x_j
		\label{eq:rolling_mean}
	\end{equation}
	where
	\begin{equation}
		w_k = \min(k,w).
		\label{eq5}
	\end{equation}
	Here, \(w\) denotes the maximum window size. At the beginning of the measurement series, the window is smaller than \(w\), since not enough previous measurements are available yet. This ensures that a defined smoothed value is obtained from the first measurement point onward. For \(k \geq w\), a full window of length \(w\) is used. 
	
	This approach preserves the complete measurement series and avoids missing initial values. At the same time, the first smoothed values are less strongly smoothed due to the smaller window size. Possible initial artifacts are therefore additionally considered later through the data-quality check and the robust auxiliary analysis.
	
	\subsection{Reference State and Degradation Index}
	TREDD does not assess degradation on the basis of absolute values, but relative to an early reference state. To this end, a baseline is determined from an early section of the smoothed measurement series. 	
	
	Let \(S_{\mathrm{ref}}\) denote the index set of the early reference segment of the smoothed measurement series.
	The reference state is given by
	\begin{equation}
		x_{\mathrm{ref}} =
		\frac{1}{|S_{\mathrm{ref}}|}
		\sum_{k \in S_{\mathrm{ref}}} \bar{x}_k .
		\label{eq:reference_state}
	\end{equation}
	This reference segment should represent an early and stable system state. Its position and length are method parameters and are specified in the experimental setup.
	
	The degradation index \(DI_k\) describes the relative deviation of the smoothed condition value from the baseline in the degradation-relevant direction:
	\begin{equation}
		DI_k =
		s \cdot
		\frac{\bar{x}_k - x_{\mathrm{ref}}}{x_{\mathrm{ref}}}
		\cdot 100
		\label{eq:degradation_index}
	\end{equation}
	
	The factor \(s\) defines the direction in which a change of the monitored condition variable is interpreted as degradation:
	\begin{equation}
		s =
		\begin{cases}
			+1, & \text{if increasing values indicate degradation},\\
			-1, & \text{if decreasing values indicate degradation}.
		\end{cases}
		\label{eq:degradation_direction}
	\end{equation}
	
	The direction factor maps the degradation-relevant direction of the monitored variable to a positive degradation index. It is therefore used instead of an absolute value, because an absolute value would also treat recovery-like or opposite-direction deviations from the baseline as degradation.
	
	For battery aging, \(s=-1\), as a decreasing capacity represents a deterioration in the battery condition. Through this direction-dependent formulation, the same algorithm can also be applied to other technical systems in which degradation is reflected by increasing or decreasing measured values.
	
	\subsection{Long-Term Trend and Drift Detection}
	The degradation index may show short-term fluctuations between neighboring observations. For example, a temporary change in operating or environmental conditions, such as an increased ambient temperature or a short load peak, may cause a transient deterioration of the measured condition variable without indicating persistent degradation.
	For this reason, an additional check is carried out to determine whether the relative deviation systematically develops in the direction of degradation over the long term. To this end, a linear trend function is applied to the degradation index:
	\begin{equation}
		\widehat{DI}(k) = a_{DI}k + b
		\label{eq10}
	\end{equation}
	The slope \(a_{DI}\)  describes the long-term direction of the change of the degradation index. A positive value of \(a_{DI}\) indicates that the degradation index increases over time.
	
	Although the original measurement series has already been smoothed, the degradation index may still show short-term fluctuations because it
	expresses relative deviations from the reference state. The first smoothing step reduces noise in the condition variable itself, whereas
	the second smoothing step is applied to the baseline-referenced degradation index. Using the same window size \(w\), the degradation trend is defined as
	\begin{equation}
		DT_k =
		\frac{1}{w_k}
		\sum_{j=\max\{1,k-w+1\}}^{k} DI_j
		\label{eq11}
	\end{equation}
	
	The drift start is not determined by a single threshold exceedance, but by a persistent exceedance of a threshold value. In the standard TREDD analysis, drift start is detected when the degradation index \(DI_k\) exceeds a threshold \(\theta\) for at least \(p\) consecutive observations.
	
	The first index at which this condition is met is defined as the drift start \(k_{\mathrm{d}}\):
	\begin{equation}
		k_{\mathrm{d}} =
		\min \left\{
		k \mid DI_j > \theta
		\ \text{for all}\
		j = k,\ldots,k+p-1
		\right\}.
		\label{eq:drift_start}
	\end{equation}
	
	In the robust auxiliary analysis, the same persistence criterion is applied to the robust degradation trend.

	The persistence criterion is a decision rule rather than an additional smoothing step. It requires the threshold condition to hold for \(p\) consecutive observations before a drift start is assigned. Thus, isolated threshold crossings are not sufficient to define \(k_{\mathrm{d}}\).
		
	\subsection{Degradation Magnitude and Degradation Rate}
	\label{sec:magnitude_rate}
	
	In addition to the drift start, two further metrics are calculated: the degradation magnitude and the degradation rate.
	
	The degradation magnitude describes the relative change of the condition-relevant variable over a selected evaluation interval. It therefore does not refer to the deterioration of the whole technical system, but to the observed change in the monitored condition variable. To this end, an early segment \(S\) and a late segment \(L\) of the smoothed measurement series are compared.
	
	Let \(k_{\mathrm{s}}\) and \(k_{\mathrm{e}}\) denote the start and end positions of the evaluation interval, and let \(m\) denote the segment length. The early and late segments are defined as
	\begin{equation}
		S = \{k_{\mathrm{s}}, \ldots, k_{\mathrm{s}}+m-1\}, \quad
		L = \{k_{\mathrm{e}}-m+1, \ldots, k_{\mathrm{e}}\}.
		\label{eq:segments}
	\end{equation}
	
	The corresponding mean values of the smoothed condition variable in these segments are
	\begin{equation}
		x_{\mathrm{start}} =
		\frac{1}{|S|}
		\sum_{k \in S} \bar{x}_k,
		\quad
		x_{\mathrm{end}} =
		\frac{1}{|L|}
		\sum_{k \in L} \bar{x}_k .
		\label{eq:xstart_xend}
	\end{equation}
	
	The relative degradation magnitude is then given by
	\begin{equation}
		D_{\mathrm{mag}} =
		s \cdot
		\frac{x_{\mathrm{end}} - x_{\mathrm{start}}}{x_{\mathrm{start}}}
		\cdot 100 .
		\label{eq:degradation_magnitude}
	\end{equation}
	
	While the degradation magnitude describes the relative change over the selected evaluation interval, the degradation rate describes the average increase of the degradation index over this interval. Using the same evaluation interval, the cycle-based degradation rate is computed as
	\begin{equation}
		r_{\mathrm{deg}}^{\mathrm{cycle}} =
		\frac{DI_{k_{\mathrm{e}}} - DI_{k_{\mathrm{s}}}}
		{k_{\mathrm{e}} - k_{\mathrm{s}}}.
		\label{eq:degradation_rate_cycle}
	\end{equation}
	
	The degradation rate is particularly relevant for comparing degradation trajectories, as it provides an estimate of the average speed at which the monitored condition variable deteriorates over the selected evaluation interval.
	
	In summary, TREDD provides several interpretable metrics: the degradation index, the long-term trend, the drift start, the degradation magnitude, and the degradation rate. These metrics form the basis for the subsequent assessment of data quality, context, and robustness of interpretation.
	
    \section{Robustness and Context Interpretation}
    \label{sec:robustness}
    The metrics calculated in Section~\ref{sec:tredd} indicate whether a persistent, baseline-referenced condition change is present. However, such a condition change is not automatically equivalent to physical aging. Measurement artifacts, unstable initial regions, changes in operating conditions, or individual outliers can produce similar trend patterns. TREDD therefore supplements the computational core algorithm with a third interpretation layer that takes data quality, context, and robust auxiliary analysis into account.
    
    \subsection{Data Quality and Context}
    The data-quality check assesses whether the measurement series allows for a reliable interpretation. In particular, it examines initial jumps,
    early instability, weak net changes over the evaluation interval, and recovery-like behavior.
    
    Let \(S_0\) denote the early segment of the measurement series used for the data-quality check. The mean level and standard deviation of this segment are defined as
    
    \begin{equation}
    	\mu_{S_0} =
    	\frac{1}{|S_0|}
    	\sum_{k \in S_0} x_k .
    	\label{eq:mu_s0}
    \end{equation}
    
    \begin{equation}
    	\sigma_{S_0} =
    	\sqrt{
    		\frac{1}{|S_0|-1}
    		\sum_{k \in S_0}
    		\left(x_k-\mu_{S_0}\right)^2
    	}.
    	\label{eq:sigma_s0}
    \end{equation}
    
    An initial jump is assessed by comparing the first measured value with the mean level of this early segment:
    \begin{equation}
    	J_{\mathrm{init}} =
    	\frac{|\mu_{S_0}-x_1|}{|x_1|}
    	\cdot 100 .
    	\label{eq:j_init}
    \end{equation}
    
    In addition, early instability is assessed via the relative dispersion of the  early segment:
    \begin{equation}
    	CV_{\mathrm{early}} =
    	\frac{\sigma_{S_0}}{\mu_{S_0}}
    	\cdot 100 .
    	\label{eq:cv_early}
    \end{equation}
    
    High values of these indicators suggest that the early part of the measurement series may not represent a stable condition level.
    
    Such measurement series are not discarded, but are treated with reduced interpretive confidence. In addition, TREDD checks whether relevant contextual variables such as temperature, current, voltage, load, or discharge duration remain stable over the evaluation interval. To this end, early and late segments of the respective contextual variables are compared.
    
    If the contextual conditions remain largely stable, an observed deterioration in the monitored condition variable can be interpreted more plausibly as aging. If, however, relevant operating or environmental conditions change at the same time, the observed condition change may be partly explained by the context. The context check therefore does not prove causality, but supports the plausibility of the degradation interpretation.
	
	\subsection{Robust Auxiliary Analysis and Confidence}
	In addition to the standard analysis, a robust auxiliary analysis is carried out. It reduces the influence of individual outliers and unstable initial values. For condition variables where decreasing values indicate degradation, such as battery capacity, a robust reference level is determined from the early reference segment, for example using an upper quantile:
	\begin{equation}
		x_{\mathrm{ref}}^{\mathrm{rob}} =
		Q_{0.9}(x_k \mid k \in S_0)
		\label{eq20}
	\end{equation}
	In addition, a rolling median can be used instead of the rolling mean. On this basis, a robust degradation index and a robust degradation trend are calculated analogously to the standard analysis. If the standard analysis and the robust analysis agree, this increases the interpretive confidence. If they differ significantly, the interpretation is assessed as uncertain or atypical.
	
	The final TREDD assessment combines drift detection, degradation magnitude, data quality, context, and robust auxiliary analysis. In the evaluation, confidence is assigned using a rule-based scheme. The term confidence refers to the rule-based strength of interpretation rather than to a statistically calibrated confidence probability.
	
	High confidence is assigned if the trajectory is not flagged as atypical, the robust slope indicates decreasing capacity, and the robust degradation magnitude is at least 10\%. Medium confidence is assigned if the trajectory is not flagged as atypical, the robust slope indicates decreasing capacity, and the robust degradation magnitude is between 5\% and 10\%. Low confidence is assigned if the degradation magnitude is below 5\%, if the trajectory shows atypical data-quality behavior, or if the robust analysis does not support a consistent degradation trend. 
	
	TREDD therefore does not produce a simple yes-or-no decision, but rather a transparent, confidence-based interpretation.
	
    \section{Experimental Evaluation}
	The publicly available NASA Lithium-Ion Battery Aging Dataset is used for the experimental evaluation~\cite{NASA_Battery_Dataset}. The dataset contains charge, discharge, and impedance measurements of several lithium-ion cells over repeated test cycles. For TREDD, only discharge cycles are considered, as capacity is available for these cycles and serves as the direct condition- and aging-relevant variable~\cite{NASA_Battery_Dataset,Berecibar2016SOH}. In the TREDD analysis, a data point corresponds to a complete discharge cycle with its associated capacity, rather than to a single voltage, current, or temperature measurement within the cycle. This results in a cyclically ordered capacity sequence \(x_1,\ldots,x_n\) for each battery. Intra-cycle measurements such as voltage, current, temperature, and discharge duration are additionally used as contextual information.
	
	\subsection{Dataset and Evaluation Setup}
	TREDD is applied to the discharge series of the batteries under consideration. For each battery, the discharge cycles are ordered, the capacity series is smoothed, and an early reference segment is used to establish the baseline. The degradation index, long-term trend, drift start, degradation magnitude, and degradation rate are then calculated. In addition, the checks described in Section~\ref{sec:robustness} regarding data quality, context, and robust auxiliary analysis are carried out.

	For the reported case study, all standard TREDD quantities are computed over the consecutive discharge-observation index \(k=1,\ldots,n\). The original NASA test ID is retained only for reporting drift locations and for the x-axis in Fig.~\ref{fig:tredd_aging}. For the standard analysis, the rolling-window size was set to \(w=10\), the reference segment contained five smoothed values starting at approximately 10\% of the discharge series, the drift threshold was set to \(\theta=10\%\), and the persistence length was set to \(p=5\) consecutive observations.
	
	In the experimental evaluation, \(k_{\mathrm{s}}\) and \(k_{\mathrm{e}}\) correspond to the 10\% and 90\% positions of the consecutive discharge-observation sequence, and the segment length is set to \(m=5\). The degradation magnitude was computed by comparing the early segment \(S\) with the late segment \(L\), as defined in Section~\ref{sec:magnitude_rate}.
	
	The degradation rate reported in Table~\ref{tab:tredd_results} is calculated per discharge observation using the consecutive index \(k=1,\ldots,n\), whereas detected drift locations are reported using the original NASA test ID.
	
	The evaluation provides an illustrative case-based assessment of TREDD on representative battery trajectories. It aims to show how the method behaves for clear, weak, gradual, and atypical degradation patterns, rather than to provide a complete statistical validation across all cells.
	
	\subsection{Results and Interpretation}
	The results show that TREDD evaluates different degradation trends in a differentiated manner. In the case of clear trends with a continuous decline in capacity, the degradation index and trend rise consistently; the standard analysis and robust analysis are largely in agreement. Weak, gradual, or unstable trends, on the other hand, are assessed with reduced confidence, particularly where no persistent drift start is detected or where data-quality indicators point to initial artifacts.
	
	\begin{figure*}[!t]
		\centerline{\includegraphics[width=0.70\textwidth]{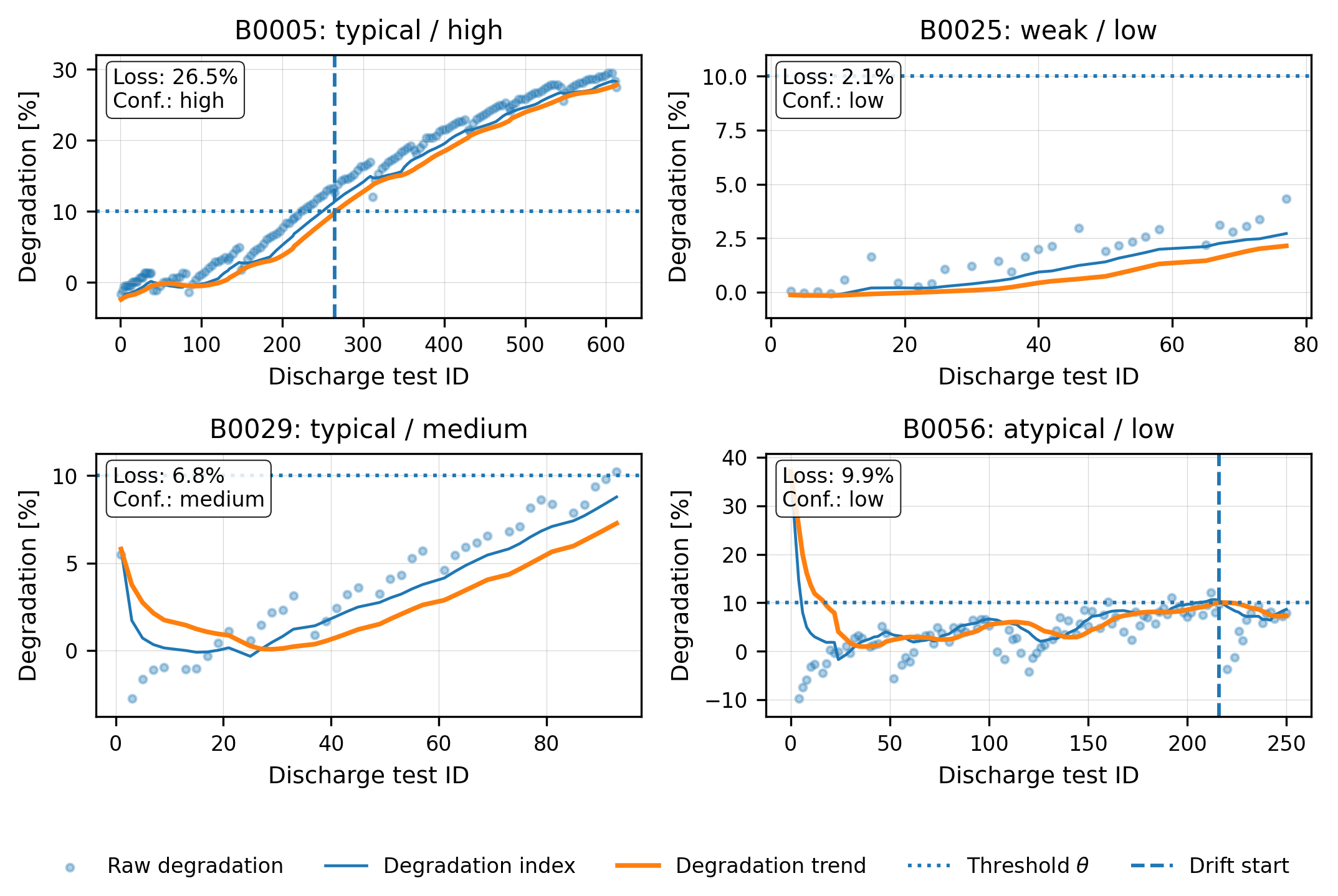}}
		\caption{TREDD diagnostic visualization for four representative NASA battery cells. Each panel shows raw baseline-referenced degradation values, the smoothed degradation index, the degradation trend, the drift threshold, and the detected drift start if present. Individual axis ranges are used because the cells contain different numbers of discharge tests and show degradation on different scales.}
		\label{fig:tredd_aging}
	\end{figure*}
		
	TREDD’s results for the NASA Lithium-Ion Battery Aging Dataset are summarized in Table~\ref{tab:tredd_results} for four representative batteries. Table~\ref{tab:tredd_results} contains the estimated degradation magnitude, the detected drift start, the degradation rate, the result of the robust analysis, the data quality, and the final confidence rating. An entry of ``n.d.'' means that no persistent drift start was detected under the selected TREDD criteria. This does not imply a lack of aging, but rather indicates that the degradation index does not exceed the threshold for the required number of consecutive observations.
	
	\begin{table}[!t]
		\caption{Representative TREDD results for selected NASA battery cells. A value of n.d. indicates that no persistent drift start was detected under the selected TREDD criteria.}
		\begin{center}
			\scriptsize
			\setlength{\tabcolsep}{2.5pt}
			\begin{tabular}{ccccccc}
				\hline
				\textbf{Cell} &
				\textbf{\begin{tabular}[c]{@{}c@{}}Capacity\\loss [\%]\end{tabular}} &
				\textbf{\begin{tabular}[c]{@{}c@{}}Drift start\\{[test ID]}\end{tabular}} &
				\textbf{\begin{tabular}[c]{@{}c@{}}Degradation rate\\{[\%/cycle]}\end{tabular}} &
				\textbf{\begin{tabular}[c]{@{}c@{}}Robust\\loss [\%]\end{tabular}} &
				\textbf{Quality} &
				\textbf{Conf.} \\
				\hline
				B0005 & 26.5 & 265  & 0.198 & 26.9 & typical  & high \\
				B0025 & 2.1  & n.d. & 0.113 & 2.5  & weak     & low \\
				B0029 & 6.8  & n.d. & 0.236 & 8.5  & typical  & medium \\
				B0056 & 9.9  & 216  & 0.121 & 7.1  & atypical & low \\
				\hline
			\end{tabular}
			\label{tab:tredd_results}
		\end{center}
	\end{table}
		
	B0005 represents a typical aging profile. The battery shows a significant capacity decline of 26.5\%, a detected drift start at discharge test ID 265, and a high confidence level. The robust analysis confirms the standard trend almost entirely, with a robust loss of 26.9\%. This case demonstrates that, for clear degradation trends, TREDD can determine both the degradation magnitude and the onset of persistent drift.
	
	B0025 represents a weak degradation trend. The capacity loss is only 2.1\%, and no drift start is detected. The low confidence level indicates that TREDD does not prematurely interpret minor changes as reliable aging.
	
	B0029 exhibits a visible aging trend but does not meet the defined persistence criterion for the drift start; the interpretation is therefore assigned medium confidence. The capacity loss is 6.8\%, while the robust analysis yields a robust loss of 8.5\%.
	
	B0056 represents an atypical trend. Although TREDD detects a capacity decrease of 9.9\% and a drift start at discharge test ID 216, the interpretation is assigned only a low confidence level. The reason lies in the data quality: the measurement series exhibits a conspicuous initial section and unstable early capacity values. The robust analysis reduces the estimated degradation magnitude to 7.1\%, which shows that the standard trend may be partially influenced by initial artifacts.

	Fig.~\ref{fig:tredd_aging} provides a diagnostic visualization of the TREDD results for the four selected batteries. Each subfigure shows the raw baseline-referenced degradation values, the smoothed degradation index, the degradation trend, the drift threshold, and the detected drift start if present. In this representation, the early reference state corresponds approximately to 0\% degradation, while increasing values indicate stronger degradation.
			
	Overall, Table~\ref{tab:tredd_results} and Fig. ~\ref{fig:tredd_aging} show that TREDD not only quantifies degradation but also assesses its interpretability. A comparison with conventional SOH- or threshold-based battery assessment approaches is left for future work, since the present study focuses on the interpretable and confidence-aware degradation assessment provided by TREDD and on the distinction between detected trends and degradation interpretations with different confidence levels.
	The absence of a drift start does not imply the absence of aging, but indicates that the defined drift condition was not met. In this way, TREDD avoids an overly confident interpretation of weak, gradual, or atypical trends.

    \section{Conclusion}
    
    This paper introduced Trend-based Reference Evaluation for Degradation Detection (TREDD), an interpretable method for detecting degradation as a persistent, trend-based deviation from an early reference state. TREDD combines rolling-window smoothing, baseline estimation, a direction-dependent degradation index, long-term trend extraction, and persistent drift detection. In addition, the method separates computational drift detection from the interpretation of drift as plausible physical aging by incorporating data-quality assessment, context checking, and robust auxiliary analysis.
    
    The approach was evaluated on the NASA Lithium-Ion Battery Aging Dataset using discharge-cycle capacity as the degradation-relevant condition variable. The results show that TREDD identifies clear degradation trajectories and assigns reduced confidence to weak, gradual, or atypical trends. In this way, the method avoids interpreting every detected or visible trend as reliable aging and instead provides a transparent, confidence-based assessment of degradation.
    
    TREDD is not intended to replace predictive models for remaining useful life or state-of-health estimation. Rather, it complements such approaches by providing an interpretable trend-based layer for degradation detection and assessment. Future work will apply TREDD to additional technical domains, integrate richer context information, and include broader cross-cell evaluation, comparison with conventional SOH- or threshold-based battery assessment approaches, statistical change-detection baselines, and parameter-sensitivity analysis.
    
    \section*{Acknowledgment}
    The authors used OpenAI's ChatGPT for grammar checking, language correction, and reference formatting. The scientific content, data analysis, results, and conclusions were developed and verified by the authors.

\end{document}